\documentclass[aps,amsmath,amsthm,amssymb,twocolumn, superscriptaddress]{revtex4-2}

\usepackage{graphicx}
\usepackage{dcolumn}
\usepackage{bm}
\usepackage[export]{adjustbox}
\usepackage[utf8]{inputenc}
\usepackage{mathptmx}
\usepackage{upgreek}
\usepackage{xfrac}
\usepackage{amsmath}
\usepackage{epstopdf}
\usepackage{hyperref}

\usepackage{xcolor}

\begin{document} 
\title{Wafer-scale fabrication of mesoporous silicon\\ functionalized with electrically conductive polymers}

\author{Manfred May}
\altaffiliation{manfred.may@tuhh.de}
\author{Mathis Boderius}

\affiliation{Hamburg University of Technology, Institute for Materials and X-Ray Physics, Denickestr. 15, 21073 Hamburg, Germany}
\affiliation{Centre for X-Ray and Nano Science CXNS, Deutsches Elektronen-Synchrotron DESY, Notkestr. 85, 22607 Hamburg, Germany}

\author{Natalia Gostkowska-Lekner}
\affiliation{Helmholtz-Zentrum Berlin für Materialien und Energie GmbH, Hahn-Meitner Platz 1, D-14109 Berlin, Germany}
\affiliation{Institut für Physik und Astronomie, Universität Potsdam, Karl-Liebknecht-Str. 24-25, D-14476 Potsdam, Germany}
\author{Mark Busch}
\affiliation{Hamburg University of Technology, Institute for Materials and X-Ray Physics, Denickestr. 15, 21073 Hamburg, Germany}
\affiliation{Centre for X-Ray and Nano Science CXNS, Deutsches Elektronen-Synchrotron DESY, Notkestr. 85, 22607 Hamburg, Germany}

\author{Klaus Habicht}
\affiliation{Helmholtz-Zentrum Berlin für Materialien und Energie GmbH, Hahn-Meitner Platz 1, D-14109 Berlin, Germany}
\affiliation{Institut für Physik und Astronomie, Universität Potsdam, Karl-Liebknecht-Str. 24-25, D-14476 Potsdam, Germany}

\author{Tommy Hofmann}
\affiliation{Helmholtz-Zentrum Berlin für Materialien und Energie GmbH, Hahn-Meitner Platz 1, D-14109 Berlin, Germany}

\author{Patrick Huber}
\altaffiliation{patrick.huber@tuhh.de }
\affiliation{Hamburg University of Technology, Institute for Materials and X-Ray Physics, Denickestr. 15, 21073 Hamburg, Germany}
\affiliation{Centre for X-Ray and Nano Science CXNS, Deutsches Elektronen-Synchrotron DESY, Notkestr. 85, 22607 Hamburg, Germany}

\begin{abstract}	
The fabrication of hybrid materials consisting of nanoporous hosts with conductive polymers is a challenging task, since the extreme spatial confinement often conflicts with the stringent physico-chemical requirements for polymerization of organic constituents. Here, several low-threshold and scalable synthesis routes for such hybrids are presented. First, the electrochemical synthesis of composites based on mesoporous silicon (pore size of 7~nm) and the polymers PANI, PPy and PEDOT is discussed and validated by scanning electron microscopy (SEM) and energy-dispersive X-ray spectroscopy (EDX). Polymer filling degrees of $ \geq \ $74 \% are achieved. Second, the production of PEDOT/pSi hybrids, based on the solid-state polymerization (SSP) of DBEDOT to PEDOT is shown. The resulting amorphous structure of the nanopore-embedded PEDOT is investigated via in-situ synchrotron-based X-ray scattering. In addition, a twofold increase in the electrical conductivity of the hybrid compared to the porous silicon host is shown, making this system particularly promising for thermoelectric applications.    
\end{abstract}	
\maketitle
\section*{Introduction}

Since the first publications on the remarkable properties of mesoporous silicon \cite{canham_silicon_1990, Lehmann1991} a lot of applications have been developed based on this material. Mesoporous silicon is used in photonics \cite{ahmed_ultra-high_2019,hernandez-montelongo_nanostructured_2015, Huber2020}, opto-fluidics \cite{Cencha2020, Dittrich2024}, gas-sensing \cite{hutter_non-imaging_2010}, biomedicine \cite{li_tailoring_2018}, energy conversion \cite{lingaraja_experimental_2023}, electromechanical actuation \cite{Brinker2020, Brinker2022, Brinker2022a} and as an anode material in batteries \cite{jia_hierarchical_2020}. It can also be used as a hard nanoporous scaffold for the study of confinement effects on the structure and dynamics of soft condensed matter systems \cite{Alba-Simionesco2006, Henschel2007, Henschel2008, Hofmann2012, Kusmin2010, Kusmin2010a, Calus2012, Vincent2017, Hofmann2017, hofmann_phonons_2021}. Conductive polymers in itself are also an enabling material class due to their general low cost and accessibility as well as their unique chemical and mechanical properties resulting in a plethora of different applications.\cite{lu_conjugated_2021} 
\\
In the last decades an increasing amount of publications in the field of inorganic-organic hybrid production has arisen.\cite{mir_revieworganic-inorganic_2018} Often these hybrids are discussed as layered systems \cite{wright_organicinorganic_2012} or as systems in which the organic constitutes the matrix and inorganics are added as additives. For example in the realm of thermoelectric hybrid materials the addition of a poly(3,4-ethylenedioxythiophene) (PEDOT) matrix to CNTs \cite{lee_improving_2016,zhou_high-performance_2021} and silicon nanowires \cite{zhang_thermoelectric_2016} leads to an increase of their thermoelectric performance factor $zT$ to a maximum of $0.44$ at room temperature. In other studies, an inorganic porous host, specifically nanoporous Au \cite{stenner_piezoelectric_2016} and mesoporous silcon \cite{Brinker2020, Brinker2022a}, are filled with the conductive polymer polypyrrole (PPy) and used as actuators and sensors. One of the main advantages of working with porous silicon as an inorganic host material is the already well-established semiconductor industry – possible meaningful applications can easily be scaled up. In the scope as a thermoelectric material, mesoporous silicon (pSi) itself shows already a small increase in the performance factor $zT$ if compared to bulk silicon. Whereas bulk Si has low $zT$ of around \(zT_{\text{293K}} < 0.01\), it can be increased upon porosification up to \(zT_{\text{293K}} < 0.02\) in accordance to the phonon-glass, electron-crystal approach.\cite{beekman_better_2015,boor_thermoelectric_2012} One of the highest \(zT_{\text{200K}} \approx 1\) for silicon was reported for a single silicon nanowire (NW)\cite{boukai_silicon_2008}.
\\
In order to increase the performance factor $zT$ a filling of the pSi host with various conductive polymers was envisioned. In first studies, an increase in conductivity from \(\sigma_{\text{293K, pSi}} = 10^{-4}\, \text{S cm}^{-1}\) to \(\sigma_{\text{300K, hybrid}} = 13\, \text{S cm}^{-1}\) could be achieved upon filling the pSi host via melt infiltration with P3HT and subsequent doping \cite{gostkowska-lekner_synthesis_2022}. However, filling degrees were restricted to approximately $50\ \%$, prompting consideration of alternative filling methods and conductive polymers. For the application as a thermoelectric material PEDOT is by far the most promising conductive polymer, with approximated $zT$ of $ > 1$ \cite{park_flexible_2013}, when prepared via chemical oxidative polymerization. In principal these electrical properties could be further increased upon confinement, because it is known that polymers tend to collectively orient in confined space, which is obviously related to their transport properties \cite{das_confinement_2022}. The direct electropolymerization of polypyrrole \cite{harraz_electrochemical_2006,brinker_giant_2020} and polybithiophen \cite{schultze_regular_1995} in pSi is already known from literature. In these studies, the mesoporous pore space was filled via the galvanostatic polymerization method. The resulting voltage-time transients feature clearly distinguishable areas from which one can interpret and track the complete filling of the pSi matrix. The electrochemical coating of a NW-forest with PEDOT is also reported, here a pulsed potentiostatic method was chosen \cite{zhu_simple_2015}. It turns out, that the filling of nanometer sized channels with conductive polymers up to sufficient depths and filling degrees is a non-trivial task. In the subsequent paragraphs, we will discuss the production of hybrids based on pSi with PPy, PEDOT and polyaniline (PANI) using electrochemical synthesis routes. Additionally we will briefly touch upon the chemical oxidative polymerization of PEDOT and lastly the solid-state polymerization (SSP) of Dibromo-3,4-ethylenedioxythiophene (DBEDOT) to PEDOT.

\section*{Materials and Methods} \label{2}
\subsection*{Mesoporous silicon fabrication} \label{2.1}
Mesoporous silicon is produced by an electrochemical etching procedure based on the anodic silicon dissolution in hydrofluoric acids. The etching solution consists of a volumetric mixture of 3:2 HF ($48\ \%$) to Ethanol (EtOH, $99\ \%$). A platin mesh works as the counter electrode (CE), the working electrodes (WE) are boron doped p+ wafers $<$100$>$ with a resistivity of 0.01 to 0.02 $\Omega$cm and a thickness of 525~µm. The samples used in the eltectrochemical polyermization procedures are etched at a current density of 12.5 mAcm$^{-2}$ for 720 s, resulting in a mesoporous silicon layer with the long pore axis normal to the sample surface. The pSi layer is still attached to the underlying bulk silicon - hereinafter referred to as an epilayer (see Fig.\ref{SEMPEDOT}).
The samples prepared for the hybrid fabrication based on the SSP, as well as the subsequent X-ray diffraction (XRD) measurements at DESY, were prepared in a different way. The used substrates are 100~µm thick, double side polished p+ wafers, with the same dopant, resistivity and orientation as the thicker wafers described before, meaning the anodic electrochemical dissolution and resulting mesoporous layer structure stays similar. The wafers were etched from both sides for 3200~s (at 12.5 mAcm$^{-2}$) creating a sandwich structure (see Fig. \ref{SEMSSP}). Naturally this is a two-step process. After each etching step every sample is cleaned in deionized water, ethanol and dried. This preparation route was necessary to ensure that a maximum pore space could be used without significant bending of the hybrid samples. Free standing membranes (detached from the underlying bulk Si) tend to crack during the recrystallization of the DBEDOT monomer after infiltration, meanwhile epilayers bend.

\subsection*{Electrochemical polymerization} \label{2.2}

All electrochemical polymerization experiments are conducted using a custom-built teflon-based polymerization cell. The cell consists of a platinum mesh counter electrode (CE), an Ag/AgCl reference electrode (RE, Sensortechnik Meisenberg), and the synthesized epilayers serve as the working electrodes (WE). All electrodes are connected to a potentiostat (Metrohm-Autolab PGSTAT 30). The polymers PPy and PEDOT are synthesized out of a solution of 0.1 M monomer and 0.1 M LiClO$_{4}$ in acetonitrile (ACN). For the galvanostatic production of the PEDOT/PPy/pSi hybrids a current of 0.255 mAcm$^{-2}$ is applied. For detailed synthesis parameters of the PANI polymer, please refer to table \ref{table1} in the results section, which provides information on the synthesis batches various compositions. When applicable, data is normalized to the substrate area in contact with the solution, without considering the increased surface area due to the porosification of the silicon. This approach was chosen due to the unknown active surface area of the pSi and the assumption that the pore ends are the preferred sites of polymerization initiation \cite{harraz_electrochemical_2006,brinker_giant_2020,schultze_regular_1995}.

\subsection*{Solid State Polymerized-PEDOT/pSi hybrid synthesis} \label{2.3}

The sandwich samples are filled with the commercially available monomer DBEDOT through melt infiltration on a hot plate at approximately 100~°C. This process is swift, with complete filling of e.g. 200~µm mesoporous silicon occurring in less than 10 seconds. Excess monomer can be easily removed using a tissue while the sample remains hot and the DBEDOT is in a molten state. Subsequently, the samples are polymerized for 96 hours at 60 °C and 10 mbar. Finally, the samples are rinsed in ethanol to remove any impurities or DBEDOT residues.

\subsection*{X-ray diffraction experiments} \label{2.4}

The powder diffraction (PXRD) samples (Fig. \ref{XRD}a.) are measured at a Bruker D8 Advance device with copper radiation of 0.154 nm wavelength in the $\Theta-\Theta$ geometry. The shown powder samples are polymerized and cleaned analogous to the procedure described in the previous section.
\\
The synchrotron-based scattering experiments conducted at the beamline P08 of PETRA III (DESY) on the sandwich SSP PEDOT/pSi hybrids are depicted in Fig. \ref{XRD}b. The incoming beam (red arrow, E = 25 keV) hits the sample and is scattered onto the 2D detector. To check whether preferred orientation of the organic species in the mesoporous pore space occurs the sample can be rotated from $\omega = 0-90 $ °. At $\omega = 0 $ ° the pore axis are parallel to the incoming beam, at $\omega = 90 $ ° perpendicular. For peak fitting (Fig. \ref{XRD}c. inset)  the free software LIPRAS (North Carolina State University) is used. Typically a Pseudo-Voigt fit gives the best result.
The samples used for the in-situ measurements during SPP of DBEDOT to PEDOT in mesoporous silicon are freshly prepared onsite in order to inhibit polymerization at room temperature. The experiments are conducted in a homemade sample cell under nitrogen atmosphere at 60 °C. 
Initially the structure of the post-imbibition DBEDOT is examined using an $\omega$ scan ranging from 0 ° to 90 ° in 1° increments. Subsequently, the temperature is raised to 60 °C, and the $\omega$ scans are repeated at intervals of 30 minutes, while simultaneously adjusting the z position each time, to mitigate beam damage.  

\subsection*{Electrical transport} \label{2.5}
The electrical conductivity is measured using a commercially available four-probe measuring device, SBA 458 (NETZSCH Gerätebau). The measurements are performed within the temperature range of 20 - 200 °C. The in-line electrical contacts are carefully positioned on the surface of the cleaned SSP-PEDOT/pSi hybrids, and a maximum current of 1 mA is injected. The samples are then stored under an inert atmosphere and exclusively retrieved for the purpose of conducting the measurements.

\section*{Results} \label{3}
\subsection*{Mesoporous silicon} \label{3.1}
The anodic electrochemical etching of p+ silicon is a well-established procedure. For the given dopant concentration, current density and electrolyte composition the resulting average pore size is around 8 nm – 9 nm in diameter. \cite{gostkowska-lekner_synthesis_2022,thelen_laser-excited_2020}
The epilayer thickness of the pSi after anodic etching of the silicon for 720 s is 11.4 $\pm$ 0.2 µm (see Fig. \ref{SEMPEDOT}). Combined with weight-loss of 9.02 $\pm$ 0.01 mg the calculated porosity is 43.2 $\pm$ 1.4 \% inside the pSi layer, which matches well with literature data \cite{brinker_giant_2020}.
The etching of the pSi/bulk Si/pSi sandwich structure, etched from both sides for 3200 s, results in the following layer thicknesses pSi = 45.8 $\pm$ 0.5 µm and bulk Si 11.6 $\pm$ 0.2 µm (Fig. \ref{SEMSSP}). In the inset of Fig. \ref{SEMSSP} a close up of the dead-end pores is shown. The remaining bulk silicon is thick enough for uninhibited current flow, leading to a sandwich structure of constant thicknesses along the whole sample size (diameter of 3.16 cm). The measured porosity inside the first pSi layers is 50.3 $\pm$ 1.3 \%, the porosity in the second 53.1 $\pm$ 1.3 \%. The increase in porosity from 43 \% to 50 \% with increasing etching time is to be expected, due to chemical etching, diffusion processes and changes in HF concentration involved during the anodic etching of silicon \cite{kumar_effect_2007}.

\subsection*{Electrochemical polymerization of PEDOT/PPy/pSi hybrids} \label{3.2}
\begin{figure}[t!]
	\centering
	\includegraphics[width=1\linewidth]{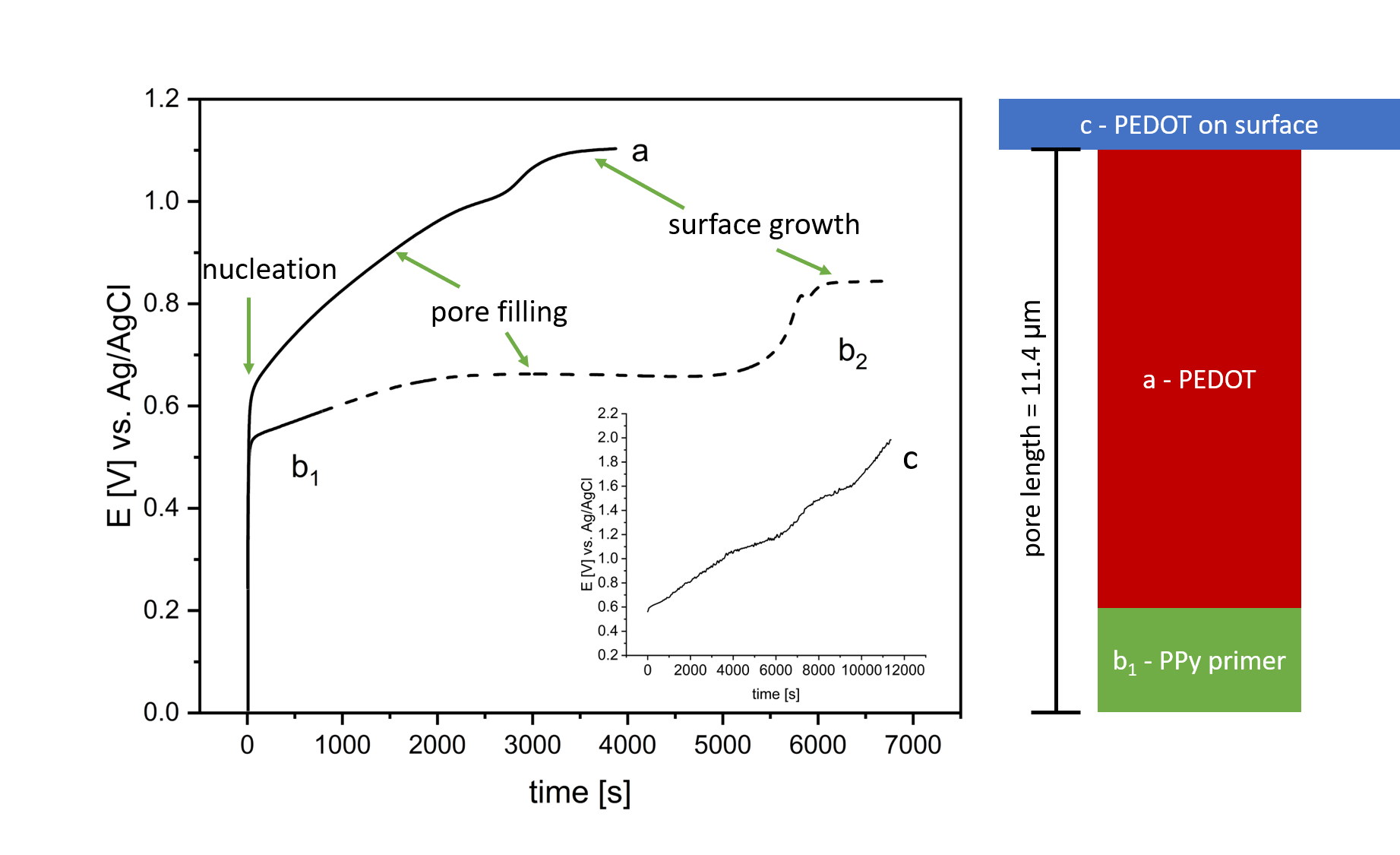}
	\caption{\textbf{Voltage-time (E-t) transients during galvanostatic deposition} of: \textbf{a} PEDOT in pSi-epilayer on a 2 µm PPy primer layer, \textbf{b$_{1}$} 2 µm PPy primer layer (for 878 s), \textbf{b$_{2}$} complete filling of the epilayer with PPy and \textbf{c} deposition of PEDOT on an epilayer without PPy primer layer. Applied current density during electropolymerization is j = 0.255 mAcm$^{-2}$. On the right side, an illustrative presentation of the filling outcomes for a single pore using different techniques labeled as \textbf{a} to \textbf{c} is depicted.}
	\label{GSPEDOT}
\end{figure}

To the best of our knowledge, this will be the first time, that direct electrochemical polyermization of PEDOT in mesoporous silicon pore space is shown. It shall be briefly mentioned that the galvanostatic deposition method on active surfaces is typically favoured compared to potentiostatic methods. This is because the necessary potential for the oxidation of the monomer and subsequent polymerization is minimized, lowering site reactions and especially the oxidation of the pSi \cite{schultze_regular_1995}.
As mentioned in the introduction, the direct electropolymerization in mesoporous silicon pore space is established for polybithiophen and PPy. Thus the expected curve shape during galvanosatic deposition is known from literature. The latter one is regulary used for the production of PPy/pSi hybrids and is shown in Fig.\ref{GSPEDOT}b$_{2}$  \cite{brinker_giant_2020}. Throughout the deposition process, there is a gradual rise in the initial voltage, reaching up to 0.5 V. This increase is associated with nucleation occurring at the bottoms of the pores and the partial oxidation of the pSi. The ensuing potential plateau is explained by the ongoing deposition of PPy inside the mesoporous silicon matrix. After around 6000 s the epilayer is completely filled and the PPy begins to polymerize at the surface. The observable increase in potential (up to 0.8 V) is ascribed to the uninhibited growth and branching of the polymer at the surface, leading to an increase in electrical resistance if compared to the directional grow inside the pSi \cite{schultze_regular_1995}.
In the inset of Fig.\ref{GSPEDOT}c the galvanostatic deposition of PEDOT on an epilayer is shown. At around 0.6 V the oxidation potential of EDOT is reached and the electropolymerization readily proceeds. Compared to the PPy deposition in Fig.\ref{GSPEDOT}b$_{2}$, the characteristic transition regimes are missing and the voltage increases linearly with time. The constant increase in voltage can be attributed to the oxidation of the pSi, meanwhile the PEDOT is polymerized only on top of the pSi, confirmed by SEM/EDX experiments. The identical problem occurred in the work of Zhu et al. during the deposition of PEDOT in/on an n-type nanowire forest \cite{zhu_simple_2015}. They circumvented the problem by thinning out the NW-forest via a KOH-tapering step and subsequent pulsed potentiostatic deposition methods. Similar approaches, both galvanostatic and potentiostatic pulsed methods were unsuccessfully tested. Additionally, to inhibit simultaneous chemical polymerization, tests were done at -18 °C -- the PEDOT still deposited always on top of the pSi. On active substrates (e.g. steel) PANI can be deposited as a corrosion protective film, by means of a PPy primer layer \cite{tuken_electrochemical_2004}. A similar approach will be pursued here but adapted for PEDOT. First an approximately 2 µm thick PPy layer ($t$ = 878 s) will be synthesized at the pore bottoms according to Fig.\ref{GSPEDOT}b$_{1}$. This PPy-primer will act as a nucleation site for the PEDOT deposition and inhibits pSi oxidation. Afterwards PEDOT is polymerized galvanostatically inside the pores on top of the PPy primer layer, the resulting voltage-time transient is shown in Fig.\ref{GSPEDOT}a. The characteristic transition regimes similar to the PPy synthesis (Fig.\ref{GSPEDOT}b$_{2}$) reappear and after around 3000 s the epilayer is completely filled, the PEDOT begins to deposit at the surface (at E = 1.1 V). Both PPy and PEDOT are deposited galvanostatically at a current density of j = 0.255 mAcm$^{-2}$ and around 2.25 electrons are needed for the polymerization of two monomer units (2) and oxidation/doping of the polymer (0.25) \cite{brinker_giant_2020,heinze_origin_2007}. Thus, the difference in deposition time, which is directly proportional to the spent charge for constant current densities, is mostly attributed to difference in molar volume of EDOT = 106.10 cm$^{3}$mol$^{-1}$ and Pyrrol = 69.16 cm$^{3}$mol$^{-1}$. Additionally, in the PEDOT/PPy/pSi hybrid less volume is available, due to the before deposited PPy primer layer. During the galvanostatic deposition of PEDOT (Fig.\ref{GSPEDOT}a) around 7.748 C of charge are used. The theoretical weight gain $w_{theo}$ can be determined using Faraday's law, expressed as $w_\text{theo} = {C\cdot M}/({z\cdot F})$, where C represents the consumed charge in coulombs (C), F is the Faraday constant in coulombs per mole (Cmol$^{-1}$), M is the formula weight of PEDOT (M = M$_{\text{EDOT}}$ – 2 protons + 0.25 $\cdot$ M$_{\text{ClO$_{4^{-}}$}}$ (gmol$^{-1}$)), and z is equal to 2.25. The calculated $w_{\text{theo}} = 5.98$  mg is significantly higher than the measured $w_{\text{meas}} = 3.53 \pm 0.01$ mg. This discrepancy can be explained by the ongoing pSi oxidation during electropolymerization, as well as delamination of PEDOT molecules. The reaction solution turns blue during ongoing polymerization. If a complete filling is assumed, the density of the deposited PEDOT inside the PEDOT/PPy/pSi hybrid is $\rho_\text{{PEDOT}} = 1.15$ $\pm$ 0.07 gcm$^{-3}$. This value is small when compared with literature bulk values $\rho_\text{{PEDOT,Lit}}$ = 1.46 gcm$^{-3}$ \cite{otero_poly_2008,delongchamp_influence_2005}, resulting in a filling degree of around 79 $\pm$ 5 \%. After rinsing with ACN no significant weight loss was measured.
\\ 
\begin{figure}[t!]
	\centering
	\includegraphics[width=1\linewidth]{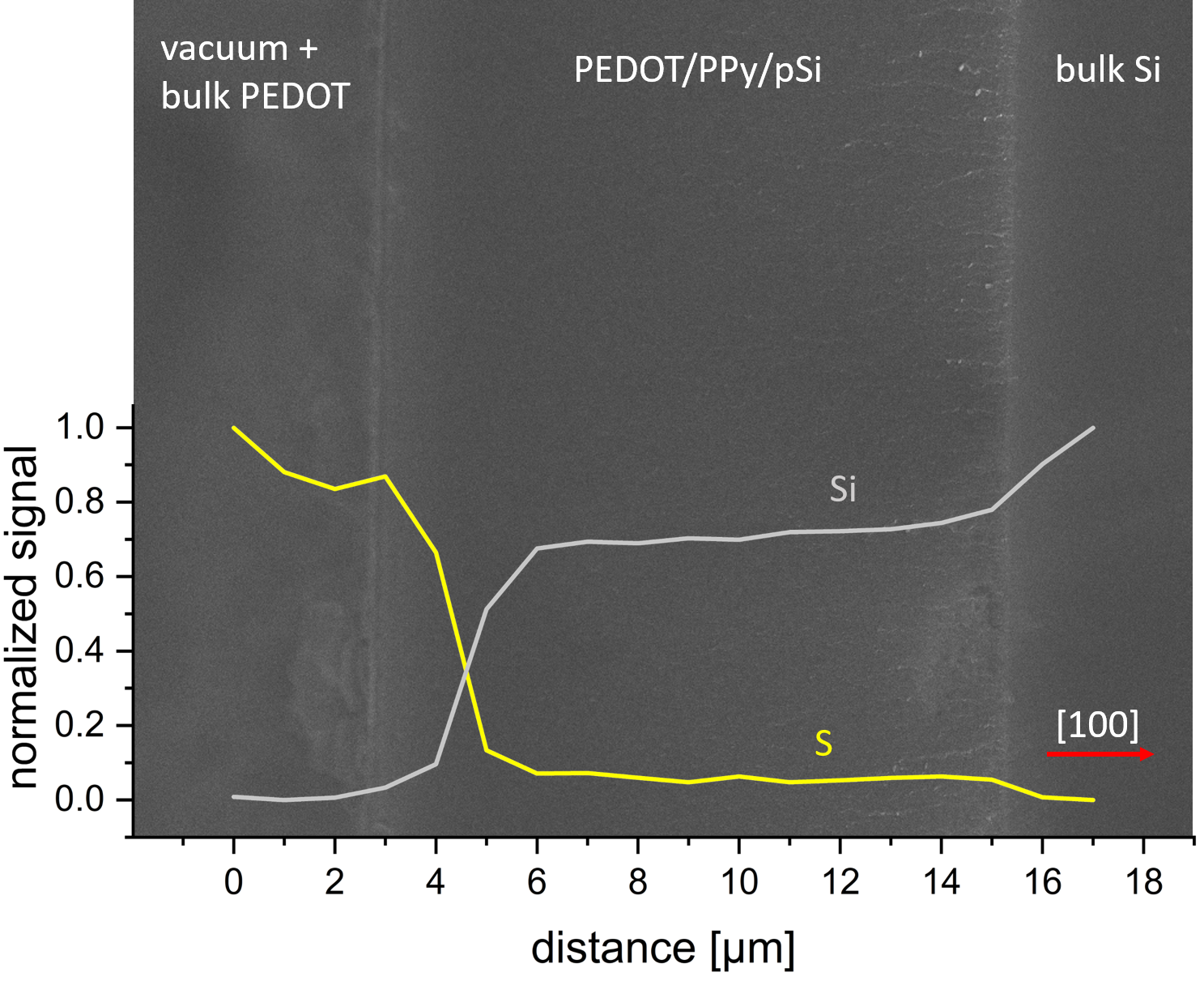}
	\caption{\textbf{SEM micrograph and superimposed silicon (grey) and sulfur (yellow) EDX signal as a funcion of distance}. In the middle part the 11.4 $\pm$ 0.2 µm thick mesoporous silicon (pSi) – filled with PPy and PEDOT is shown. The hybrid is attached to the underlying bulk Si (right site). In interest for greater clearity the EDX data for carbon, nitrogen and chloride are not shown. The red arrow indicates the crystal orientation as well as the pore axis perpenticular to the sample surface.} 
	\label{SEMPEDOT}
\end{figure}
The qualitatively filling of the synthesized PEDOT/PPy/pSi hybrid was checked via SEM/EDX measurements (Fig.\ref{SEMPEDOT}). The sulfur signal of the EDOT monomer is detectable inside the whole 11.4 $\pm$ 0.2 µm  pSi layer. Interestingly, no decrease in the sulfur signal can be seen at the pore bottom, where the PPy is deposited. This can be mostly explained by the huge interaction volume due to the necessary high acceleration voltage for the detection of sulfur during SEM/EDX experiments. This is also the reason for the high sulfur signal near the sample surface, which is attributed to bulk PEDOT at the surface.
\subsection*{Electrochemical polymerization of PANI/pSi hybrids} \label{3.3}
\begin{figure}[t!]
	\centering
	\includegraphics[width=1\linewidth]{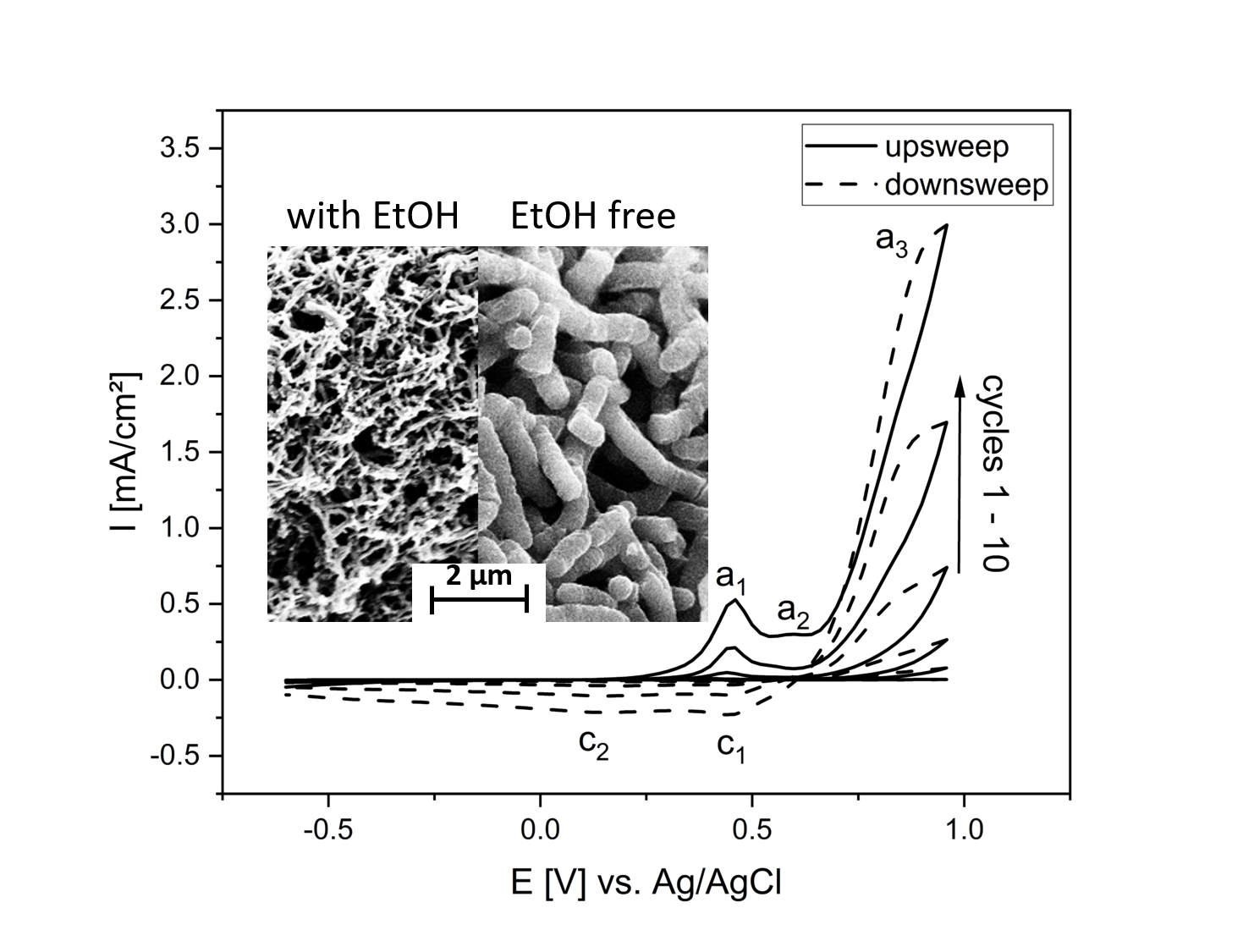}
	\caption{\textbf{Potentiodynamic synthesis of PANI} according to PANI 5, table 1 on a p+ wafer, sweep rate 20~mVs$^{-1}$. Only the first 10 cycles are shown. The inset shows SEM micrographs and the effect of the added EtOH on the PANI microstructure.} 
	\label{CV}
\end{figure}

\begin{figure}[t!]
	\centering
	\includegraphics[width=1\linewidth]{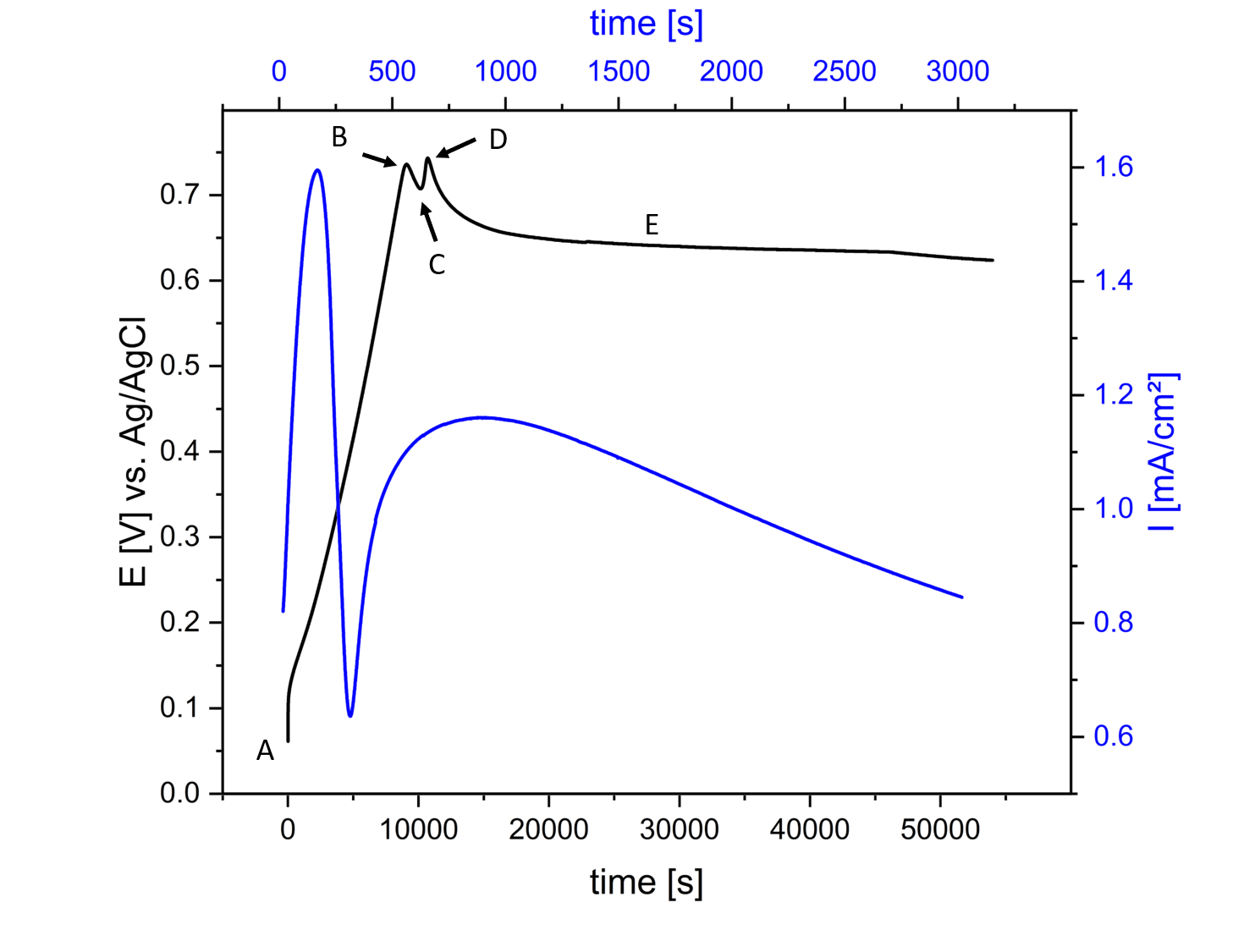}
	\caption{\textbf{Galvanostatic (black, I =  0,255 mAcm$^{-2}$) and potentiostatic (blue, E = 0,9 V) deposition of PANI 5 on pSi epilayers.} } 
	\label{GSPS}
\end{figure}

Anodically etched silicon is hydrogen terminated (hydrophobic) and PANI is typically synthesized from an acidic aqueous solution (PANI 4, table \ref{table1}), thus two different synthesis routes for the hybrid production were envisioned. First the PANI/pSi synthesis out of nonaqueous solvents was tried. Miras et al reported the successful synthesis of PANI, out of a 0.1 M aniline 0.5 M LiClO$_{4}$ in ACN (PANI 1) solution, which is: “electroactive and stable”. \cite{miras_preparation_1991} A result which could not be reproduced on any silicon surface. Only a brown, electro-inactive PANI film with low electrical conductivity was deposited, a result which is known for the synthesis of PANI in basic, neutral or weakly acidic media. \cite{de_souza_gomes_oxidative_2012} Similar results were obtained for the polymers PANI 2-3. \cite{pandey_electrochemical_2002}
\\
\begin{table}[htbp]
  \centering
  \caption{Overview of the different reaction solutions for the electrochemical synthesis of the PANI/pSi hybrids. *TBAP = tetrabutylammonium perchlorate, *TBATFB = tetrabutylammonium tetrafluoroborate}
   \begin{tabular}{lccc}
\hline
     sample & aniline [M] & solvent & electroclyte [M] \\
\hline
    PANI 1 \cite{miras_preparation_1991} & 0.1  & ACN & 0.5 LiClO$_4$  \\
PANI 2 \cite{pandey_electrochemical_2002}& 0.4 & ACN & 0.1 TBAP*  \\
PANI 3 \cite{pandey_electrochemical_2002}& 0.4 & ACN & 0.1 TBATFB*  \\
PANI 4 \cite{gvozdenovic_electrochemical_2014}& 0.25 & H$_2$O & 1 HCl  \\
PANI 5 [this work]& 0.25 & 3 M EtOH (aq) & 1 HCl   \\
\hline
    \end{tabular}%
  \label{table1}
\end{table}%
In the second approach, 3 M EtOH were added to the standard synthesis solution (PANI 4, table \ref{table1}) to guarantee a successful infiltration of the solution into the hydrophobic mesoporous pore space. The influence of the EtOH addition during potentiodynamic PANI deposition on top of a p+ wafer was tested in Fig. \ref{CV}, as well as the effect on the resulting PANI morphology towards finer structures (Fig. \ref{CV} inset), which is well known due to solvation processes of PANI molecules by EtOH \cite{yazdanpanah_synthetic_2019}. The deposited PANI 5 was green in colour (emeraldine salt) and not distinguishable by eye, or via a comparison of the cyclic voltammograms, from EtOH-free synthesized PANI 4. All peaks a$_{1}$-a$_{3}$ and c$_{1}$, c$_{2}$ are well described in the literature and are attributed to the different oxidation states of PANI, except for the anodic peak a$_{2}$, which refers to hydrolysis and degradation products. For more detailed informations regarding the peaks and factors influencing the PANI synthesis, please refer to the excellent work of Gvozdenovic et. al.\cite{gvozdenovic_electrochemical_2014}. The anodic peak a$_{3}$ is the superposition of the formation of fully oxidized pernigraniline salt and the oxidation of the silicon substrate, which is favoured at higher voltages. Furthermore, after reversing the potential scan direction the downsweep lies at higher currents than the upsweep. This trace-crossing effect is not to be confused with the so-called “nucleation loop”, which is typically oberserved during the first cycle in the electropolymerization of conductive polymers. Rather it is based on the slow formation redoxactive intermediates at the diffusion layer of the silicon (WE) and their ensuing oxidation. \cite{heinze_origin_2007} An increase in PANI layer thickness and accompanying capacitance increase with rising cycle count can be seen in the growing gap between up- and downsweep.
\\
Subsequently the PANI 5 was synthesized galvanostatically on an epilayer, the resulting potential-time transient is shown in Fig. \ref{GSPS} (black). Striking are the 9000 s it took, to form the first polymer monolayer on top of the epilayer (A-B), accompanied by simultaneous pSi oxidation. Especially when compared to the results for PPy and PEDOT (Fig. \ref{GSPEDOT}), wherein the growth stage was reached almost immediately. A stable growth stage (E) at E = 0.65 V was achieved after around 15000 s, in between B-C new nucleation sites are formed and the PANI film begins to grow (C-D), after which the potential begins to fall again.\cite{chen-yang_electropolymerization_2004} When the PANI 5 is galvanostatically synthesized at the same conditions on a p+ silicon wafer, the 2nd peak disappears and the growth stage is reached almost immediately. The rise in potential of the 2nd peak could be explained by the filling of the pores and and subsequent deposition on top of the epilayer, for which a higher potential is necessary due the uninhibited network-like grow – compared to the directional grow inside the the pores \cite{brinker_giant_2020}. But SEM/EDX studies reveal that no PANI was deposited inside the pSi layer. Additionally, no positive effect of a PPy-primer layer, identical to the one described in figure~\ref{GSPEDOT}, on the deposition of PANI into the mesoporous pore space could be observed. In order to reduce the time and pSi oxidation during the galvanostatic deposition, a necessary minimum potential of 0.9 V was applied for the potentiostatic polyermerization of PANI (Fig. \ref{GSPS} blue). During potentiostatic deposition the current-time transients for the deposition on a p+ wafer (not shown) and the epilayers are similar in shape. The initial increase can be attributed to the monomer oxidation and nucleation processes \cite{schultze_regular_1995}, after which the PANI directly polymerizes on top of the epilayer as a green film (t $>$ 250 s), which can be seen directly by eye. 
\\
When comparing the results from the successful polymerization of PEDOT/PPy/pSi hybrids with the unsuccessful pore filling during the PANI/PPy/pSi hybrid synthesis, most noteworthy is the increased oxidation potential of the aniline compared to EDOT and the accompanying increased pSi oxidation (amplified at higher voltages).

\subsection*{Solid State Polymerized-PEDOT/pSi hybrids via monomer imbibition and subsequent SSP} \label{3.4}

Due to the inevitable synchronized oxidation of the pSi during the synthesis of the PEDOT/PPy/pSi hybrids (Fig. \ref{GSPEDOT}) and the limited maximum sample thickness of the resulting hybrids to a around 10 - 20 µm, the direct oxidative chemical polyermerization of PEDOT in pSi, which promised even better electrical properties \cite{park_flexible_2013}, was tested. An approach which was quickly dropped, where even after serval iterations of cleaning and polymerization, similar to the production of PEDOT filled tantalum powders \cite{elschner_pedot_2010}, no high filling degrees could be achieved. In general, pristine conductive polymers tend to show low processability \cite{ruiz_improving_2013} and even if they are processable in molten state \cite{gostkowska-lekner_synthesis_2022}, high filling degrees in pSi are a challenging task.
\\
Hong Meng et. al found out by coincidence that after long time storage of DBEDOT at room temperature, the grey powder turned from grey into blue, conductive PEDOT (see inset Fig. \ref{XRD}a). This process named solid state polymerization (SSP) is not given much attention in the scientific community, due to the normally quite sufficient and well researched chemical and electrochemical synthesis routes. Importantly, the DBEDOT monomer has a melting point of 96 °C and nearly no polymerization occurs in molten state.\cite{meng_solid-state_2003} In the following chapters the synthesis of SSP-PEDOT/pSi hybrids will be discussed, as well as the resulting structure during confinement and the effect on the electrical transport properties of the hybrid.

\subsection*{Synthesis of  SSP-PEDOT/pSi hybrids} \label{3.4.1}

The spontaneous imbibition of the molten DBEDOT at $T \geq$ 96 °C into pSi is quite fast, 200 µm thick pSi layers are completely filled in $t <$ 10 s. Infiltration processes based on polymers are comparatively slower (e.g. $>$ 10 h \cite{gostkowska-lekner_synthesis_2022}). After infiltration the monomer is polymerized at T = 60 °C for 96 h as previously discribed. An important advantage of this synthesis route, other than the simplicity, is the ability to fill various porous structures. The filling process is based on capillary flow, eliminating the need for complex measures such as ensuring electrical contact with the porous substrate, which is essential for advanced techniques like electrochemical polymerization. Unconventional hybrid materials, e.g. like brick-PEDOT supercapacitors (originally produced via vapor phase polymerization \cite{wang_energy_2020}) should be in the scope of this technique.
\\
\begin{figure}[t!]
	\centering
	\includegraphics[width=1\linewidth]{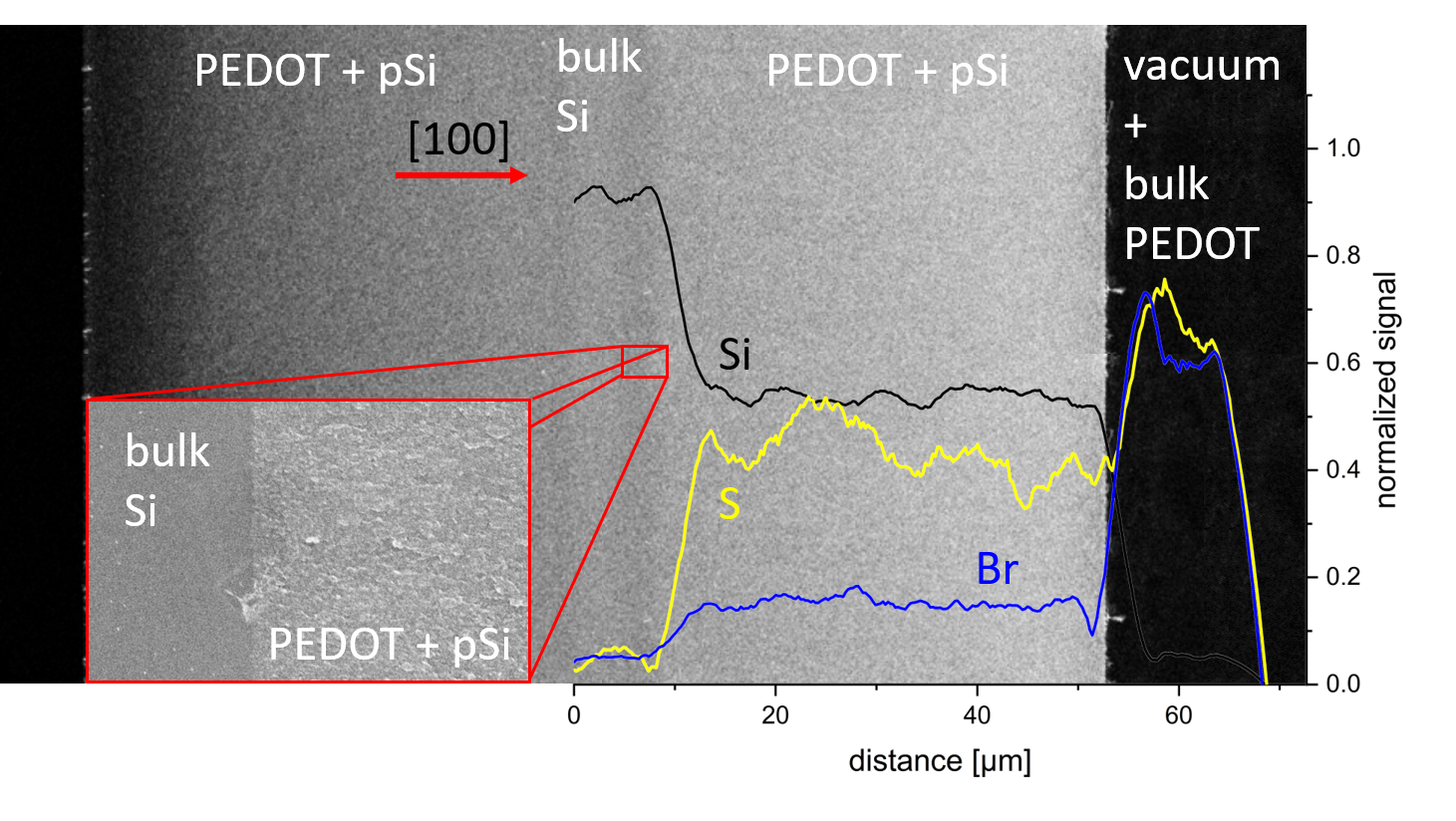}
	\caption{\textbf{SEM micrograph and superimposed silicon (black), sulfur (yellow) and bromine (blue) EDX signal as a funcion of distance}. In the middle part the 11.6 $\pm$ 0.2 µm bulk Si layer is visible, left and right the 45.8 $\pm$ 0.5 µm thick pSi layer filled with PEDOT. The close-up shows an interface (dead-end pores) between the pSi and bulk Silicon.} 
	\label{SEMSSP}
\end{figure}
In Fig. \ref{SEMSSP} the complete filling of the sandwich structure was qualitatively confirmed. During SSP 8.8 $\pm$ 1.1 wt\% is lost as Br$_2$ molecules, normalized to the DBEDOT amount. Which is less, than the expected 21.3 wt\% during SSP of bulk or powder samples, calculated from the following chemical compositions C$_6$H$_4$Br$_{1,2}$O$_2$S \cite{meng_solid-state_2003} and C$_6$H$_4$Br$_2$O$_2$S (DBEDOT). During the dimerization of two DBEDOT molecules elemental Br$_2$ is released \cite{kim_room_2012}, which in turn acts partly as the dopant (Br$_3^-$), leading to an empirical formula of  C$_6$H$_4$Br$_{1,2}$O$_2$S and a bipolaron per every five thiophene units \cite{meng_solid-state_2003}. The difference in empirical and theoretical weight-loss during SSP, can be explained by unwanted site reactions between the pSi and Br$_2$ \cite{bressers_etching_1996} in combination with possibly trapped Br$_2$.

\subsection*{X-ray scattering experiments} \label{3.4.2}
\begin{figure*}[t!]
	\centering
	\includegraphics[width=1\linewidth]{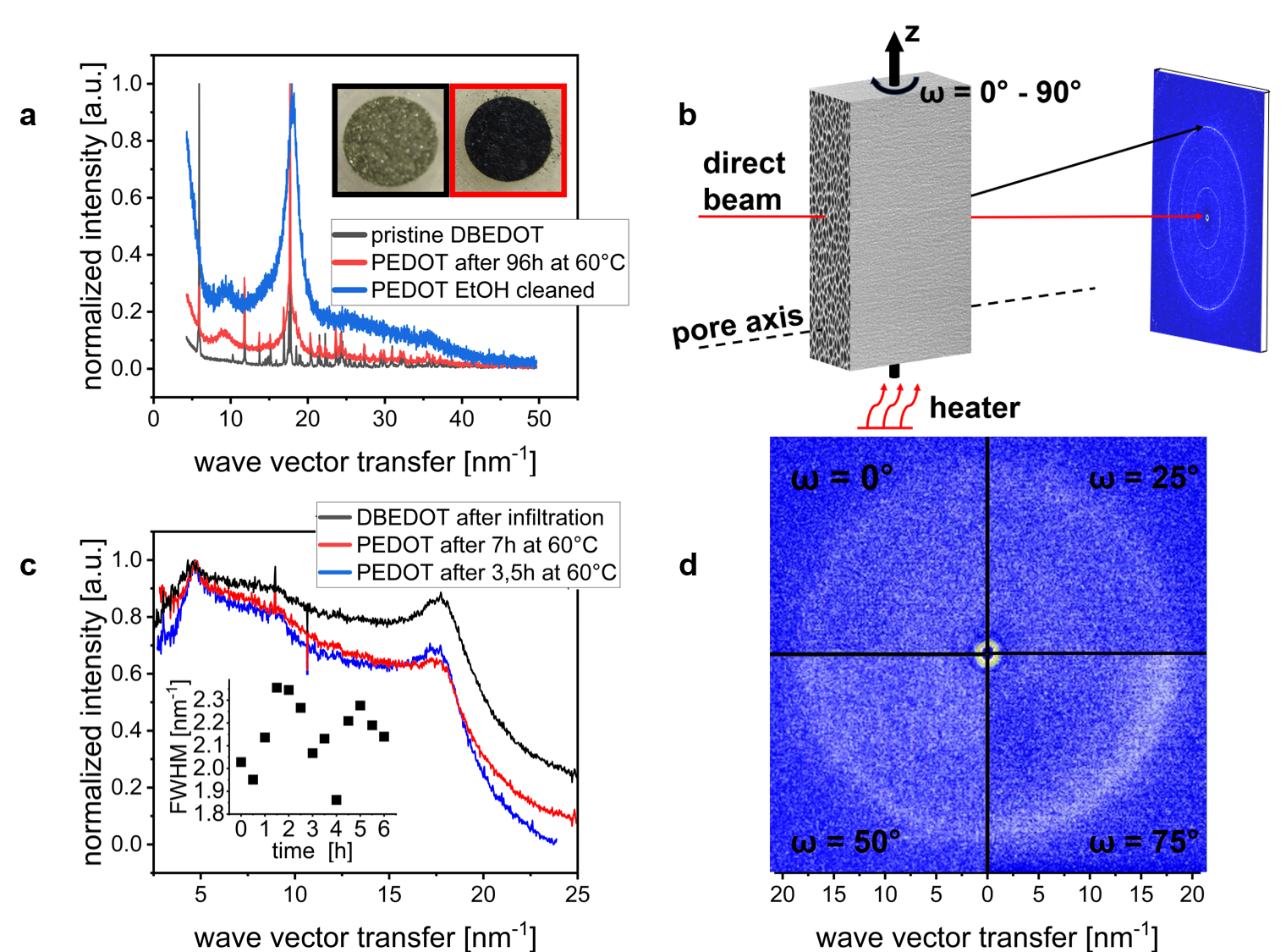}
	\caption{\textbf{X-ray scattering experiments on SPP-PEDOT/pSi hybrids.} In \textbf{a.} PXRD diffractograms (lab source) for DBEDOT, PEDOT and EtOH cleaned PEDOT samples are shown - accompanied by an increasing FWHM of the Q = 17.6 nm$^{-1}$ reflection during SSP and powder colour change from grey (DBEDOT) to blue (PEDOT). In \textbf{b.} the schematic of the X-ray experiments at P08 DESY are shown. Importantly, the angle $\omega$ describes the angle of the incoming beam (red) to the sample surface and pore direction, at $\omega$ = 0° the pores are parallel, at $\omega$ = 90° perpenticular to the beam. In \textbf{c.} the diffractograms (integrated from 2D detector images) of DBEDOT (black) directly after infiltration, and PEDOT after 3.5 h at 60 °C (blue) and 7 h at 60 °C (red) are shown. In the inset the change in FWHM of the Q = 17.6 nm$^{-1}$ reflection with reaction time is demonstrated. In \textbf{d.} a compiled 2D detektorimage (each quadrant is a cutout at different $\omega$ values) is depicted, indicating no preferred orientation of DBEDOT molecules directly after infiltration.} 
	\label{XRD}
\end{figure*}
\begin{figure*}[t!]
	\centering
	\includegraphics[width=1\linewidth]{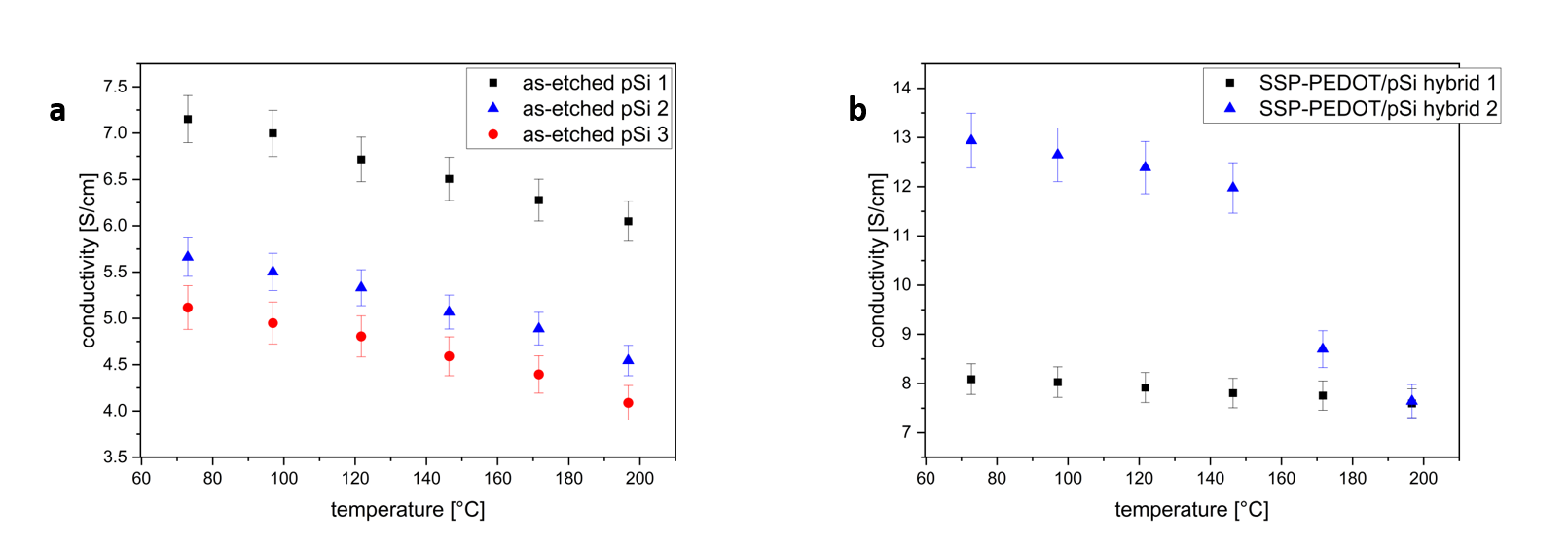}
	\caption{\textbf{Electrical conductivity of a. as-etched pSi sandwich structure and b. SSP-PEDOT/pSi hybrids as a function of temperature.} } 
	\label{electrical}
\end{figure*}

In Fig. \ref{XRD} the results of the X-Ray experiments performed on the SSP-PEDOT/pSi hybrids is shown. From the PXRD patterns in Fig. \ref{XRD}a. it is obvious that DBEDOT is a highly crystalline monomer (black) featuring numerous sharp peaks, which crystallizes in a monoclinic structure \cite{meng_solid-state_2003}. During the SSP the high crystallinity of the sample is lost (red curve) and an especially broad peak appears at Q = 17.6 nm$^{-1}$. After an EtOH rinse the resulting diffractogram is identical to previously reported XRD patterns for PEDOT \cite{han_facile_2006}, with the Q = 17.6 nm$^{-1}$ (d = 3.4 \text{\AA}, indexed as (020)) referring to the interchain planar ring-stacking distance, which is perpendicular to the conjugated polymer backbone – two crystal directions which show high electrical conductivity. \cite{aasmundtveit_structure_1999,han_facile_2006} 
\\
Directly after infiltration of the pSi with the DBEDOT, the DBEDOT precipitates in an amorphous manner (Fig. \ref{XRD}c. black) in which the long-range ordering is significantly reduced. An effect which is often found for nanoconfined materials, in particular embedded in mesoporous media, where the pure geometrical constriction and the interaction with the pore walls often hinders crystallization, leads also to substantial reduction in the freezing/melting transitions \cite{Schaefer2008,Alba-Simionesco2006,  Huber2015, meldrum_crystallization_2020} and the favorization of glassy compared to crystalline phases \cite{Henschel2010}. Importantly the DBEDOT keeps its high order of crystallinity if recrystallized as a bulk film or in macroporous silicon.
Contrary to the results for the powder samples, no increase of the FWHM (Fig. \ref{XRD}c inset) of the Q = 17.6 nm$^{-1}$ peak with ongoing SSP can be seen. The average coherence length (Scherrer equation) for the listed FWHM is $\xi$ = 2.8 $\pm$ 0.2 nm, which is smaller than the pSi pore size and further illustrates the short range ordering. Meanwhile the successful polymerization can be checked gravimetrically, via an EtOH cleaning procedures (PEDOT is insoluble, DBEDOT soluble) and conductivity measurements (Fig. \ref{electrical}). This is interesting, because Hong Meng et al. proposed a polymerization direction along monomer stacks (due to the short halogene$\cdots$halogene distance between two monomer units), which has to be accompanied by significant rotation of molecules, which should be influenced given the steric hindrances due to confinement \cite{meng_solid-state_2003}. Only weak indications of molecular texturation within the mesopores were observed. The enhanced intensity with increasing rotation angle $\omega$, illustrated in Fig. \ref{XRD}d., is primarily associated with a larger illuminated sample volume at higher $\omega$.


%
%
%
\subsection*{Electrical transport measurements}\label{3.4.3} 
In Fig. \ref{electrical} the results of the electrical conductivity measurements of the SSP-PEDOT/pSi hybrids are shown. The high conductivity ($\sigma$) of the as-etched pSi samples (Fig. \ref{electrical}a.), when compared to the electrical conductivity of free-standing pSi membranes $\sigma$ $<$ 10$^{-4}$ Scm$^{-1}$ \cite{gostkowska-lekner_synthesis_2022}, is due to the 11.6 $\pm$ 0.2 µm thick bulk Si layer inside the sandwich structure. The general trend towards lower conductivities with increasing temperatures is known as the extrinsic region in doped semiconductors. It can be explained by the reduced carrier mobility caused by thermal scattering, meanwhile the carrier density stays nearly constant in this temperature region \cite{arora_electron_1982}.
\\

The same trend is visible for the hybrid samples, indicating that the electrical transport is also governed by the bulk Si layer. Importantly the conductivity is increased up to a maximum value of ~ $\sigma$  = 13 Scm$^{-1}$, which corresponds to roughly a doubling in conductivity, which elucidates the successful SSP from DBEDOT to PEDOT during confinement. Most probable explanation for the difference between the hybrid samples is a difference in local filling degrees, due to the synthesis procedure described in the materials and methods section. 

\section*{Conclusion and outlook}
To summarize, this study introduces two innovative approaches for creating hybrids of PEDOT and mesoporous silicon. The first method involves the successful synthesis of PEDOT/PPy/pSi hybrids, achieved through the application of a PPy primer layer and subsequent galvanostatic deposition of PEDOT directly into the mesoporous pore space. The result is a hybrid with a high degree of polymer filling of 79 $\pm$ 5 \%.
\\
Secondly the hybrid synthesis based on the spontaneous imbibition of molten DBEDOT and ensuing SSP to PEDOT during confinement. In the X-ray diffraction experiments it could be shown, that the DBEDOT precipitates in an amorphous manner inside the mesoporous silicon. Thus, for the application as a thermoelectric material it would be interesting to figure out the minimal necessary pore size at which a preferred orientation occurs, which should give rise to anisotropic and possibly enhanced electrical properties \cite{das_confinement_2022}. Additionally, in order to reduce side reactions between the Br$_2$ and Si and simultaneous lower the polymerization temperature during SSP, the Br inside the DBEDOT monomer can be substituted with for example iodine \cite{meng_solid-state_2003}. The proposed synthesis route can be easily transferred to other porous systems and research fields - we hope that based on this work, SSP of PEDOT will gain more attention and appreciation in the scientific community. It has been shown that by filling the mesoporous structure with PEDOT, the conductivity can be increased by a factor of 2.
\\
Furthermore, this work provides first insights into the chemical oxidative polymerization of PEDOT in pSi as well as the unsuccessful electrochemical polymerization of PANI inside pSi. This may serve as a cautionary note for other researchers, preventing them from investing resources into a potentially challenging pursuit. While we do not assert that the electrochemical synthesis of PANI within mesoporous silicon is unattainable, our findings highlight the difficulties encountered in the hybrid synthesis process, despite a comprehensive understanding of electrochemical synthesis methods.
 
%
%
%
\medskip
\begin{acknowledgements}
This work was supported by the Deutsche Forschungsgemeinschaft (DFG) within the project "Hybrid thermoelectrics based on porous silicon: Linking Macroscopic Transport Phenomena to Microscopic Structure and Elementary Excitations" Project number 402553194 and the Collaborative Research Initiative SFB 986 “Tailor-Made Multi-Scale Materials Systems” project number 192346071. We acknowledge Deutsches Elektronen-Synchrotron DESY (Hamburg, Germany), a member of the Helmholtz Association HGF, for the provision of experimental facilities. Parts of this research were carried out at PETRA III and we would like to thank Dr. Florian Bertram for assistance in using the P08 beamline. Beamtime was allocated for proposal I-20210755.\\

\textbf{Data and materials availability}
All data that is needed to evaluate the conclusions is presented in the paper and in the supplementary materials. The electrochemical and x-ray diffraction raw data sets are available at TORE (\url{https://tore.tuhh.de/}), the Open Research Repository of Hamburg University of Technology, at the doi:\url{https://doi.org/10.15480/882.9049}.\\

\textbf{Author contributions}
M.M., M.Bo. and P.H. conceived the experiments. M.M. and M.Bo. performed the material synthesis and electrochemical data evaluation. M.M. and M.B. performed the X-ray scattering experiments at the synchrotron PETRA III at Deutsches Elektronen-Synchrotron DESY. The data analysis was performed by M.M.. The electrical conductivity measurements were performed by N. G.-L. and M.M., M.B., K.H., T.H. and P.H. wrote the manuscript. All authors revised the manuscript. The authors declare that they have no competing interests.
\end{acknowledgements}
%
%
%
\medskip
\clearpage
%
\end{document}